\documentclass[a4paper,11pt]{article}
\usepackage[margin=2.5cm]{geometry}
\usepackage{setspace,graphicx,url,hyperref,enumitem}

\usepackage{amsmath,amssymb,amsfonts,amsthm,natbib}

\begin{document}
\onehalfspacing
\title{Solving large Maximum Clique problems on a quantum annealer}
\author{Elijah Pelofske\footnote{Los Alamos National Laboratory, Los Alamos, NM 87545, USA},
Georg Hahn\footnote{Lancaster University, Lancaster LA1 4YW, U.K.}, and Hristo Djidjev\footnotemark[1]}
\date{}
\maketitle

\begin{abstract}
Commercial quantum annealers from D-Wave Systems can find high quality solutions of quadratic unconstrained binary optimization problems that can be embedded onto its hardware. However, even though such devices currently offer up to 2048 qubits, due to limitations on the connectivity of those qubits, the size of  problems that can typically be solved is rather small (around 65 variables). This limitation poses a problem for using D-Wave machines to solve application-relevant problems, which can have thousands of variables. For the important Maximum Clique problem, this article investigates methods for decomposing larger problem instances into smaller ones, which can subsequently be solved on D-Wave. During the decomposition, we aim to prune as many generated subproblems that do not contribute to the solution as possible, in order to reduce the computational complexity. The reduction methods presented in this article include upper and lower bound heuristics in conjunction with graph decomposition, vertex and edge extraction, and persistency analysis.
\end{abstract}

\section{Introduction}
\label{sec:intro}
Quantum annealers such as the commercially available ones manufactured by D-Wave \citep{D-WaveSystems2000QuantumToday} aim to solve quadratic unconstrained binary optimization (QUBO) problems by looking for a minimum-energy state of a quantum system implemented on the hardware. Notably, due to the stochastic nature of annealing, the approximations of the global minimum returned by D-Wave are not always identical, while the solving time is independent of the problem instance. The class of QUBO problems includes many well-known NP-hard problems \citep{Lucas2014}, for which no classical polynomial algorithms are known. In this article, we focus on the NP-hard problem of finding a maximum clique in a graph.

The D-Wave quantum processing unit is designed to minimize the \textit{Hamiltonian}
\begin{equation}
	H = \sum_{i \in V} a_{i} x_j + \sum_{(i,j ) \in E} a_{ij} x_i x_j,
	\label{eq:hamiltonian}
\end{equation}
where $V=\{1,\ldots,n\}$, $\{x_i~|~i\in V\}$ is a set of $n$ binary variables, and $E \subseteq V \times V$ describes quadratic interactions between them. The coefficients $a_i$ are the weights of the linear terms, and $a_{ij}$ are the weights of the quadratic terms (\textit{couplers}). If in \eqref{eq:hamiltonian} we have $x_i \in \{0,1\}$ for all $i \in V$, the Hamiltonian is called a \textit{QUBO}, whereas if $x_i \in \{-1,+1\}$ for all $i \in V$, it is called an \textit{Ising} Hamiltonian. The Hamiltonian for the maximum clique problem we consider is a QUBO.

The \textit{maximum clique problem (MaxClique)} is defined as the problem of finding a largest \textit{clique} (a completely connected subgraph) in a graph $G = (V, E)$. The \textit{clique number} is the size of the maximum clique. In \cite{Chapuis2017}, the QUBO formulation of the maximum clique problem is given as
\begin{equation}\label{eq:MC_QUBO}
	H = -A\sum_{i \in V} x_i + B\sum_{(i,j ) \in \overline{E}} x_i x_j,
\end{equation}
provided $B/A$ is sufficiently large, where the choice $A=1$ and $B=2$ ensures equivalence of the maximum clique problem and the Hamiltonian in \eqref{eq:MC_QUBO}, and $\overline{E}$ is the edge set of the complement graph of $G$.

The D-Wave annealer has a fixed number of operational qubits which are arranged in a certain graph structure, consisting of a lattice of bipartite cells, on a physical chip~\cite{Chapuis2017}. D-Wave 2X has 1152 physical qubits, while the newest D-Wave 2000Q model has up to 2048 physical qubits. For many NP-hard problems, however, the connectivity structure of a QUBO (that is, the graph defined by its non-zero coefficients $a_{ij}$) does not match the specific connectivity of the qubits on the D-Wave chip. In particular, this is the case for general instances of MaxClique. This problem can be alleviated by computing a \textit{minor embedding} of a complete graph onto the D-Wave architecture, thus guaranteeing that any QUBO connectivity can be mapped onto the D-Wave chip. In such an  embedding, a connected set of physical qubits is identified to act as one logical qubit, which further reduces the number of available qubits~\citep{TechnicalDescriptionDwave}. Even if the connectivity structure required to solve a QUBO matches the one of D-Wave, the scaling of problem instances quickly exhausts the capacity of the D-Wave chip. In both cases, classical preprocessing can attempt to split up large problem instances into smaller ones, which can be solved on D-Wave. General purpose approaches for QUBO decomposition exist \citep{Boros2002}, even though they are not guaranteed to always work for arbitrary QUBO instances.

Special purpose techniques exist for decomposing certain NP-hard problems. This paper investigates a decomposition algorithm for finding a maximum clique in a graph \citep{Chapuis2017}, where the problem reduction does not result in a loss of information regarding the maximum clique of the original problem. The decomposition algorithm works by recursively splitting a graph at each level in such a way that the maximal clique is guaranteed to be contained in either of the two generated subgraphs. To reduce the computational complexity, we attempt in this article to identify and exclude those generated subgraphs that cannot contain a maximum clique. This is achieved by computing bounds on the maximal clique size at each iteration. The purpose of this article is to investigate several such techniques with the aim of demonstrating the viability of solving larger optimization problems with D-Wave than previously possible.

Decomposition has already been suggested to solve several NP-hard problems in \cite{Tarjan1985}, and since used for NP-hard problems such as domination-type problems and graph coloring in \cite{Rao2008}. For the Maximum Independent Set problem, an equivalent formulation of MaxClique, polynomial-time algorithms are known for many special graph classes \citep{Dabrowski2011,Grotschel1988,Minty1980}, including several algorithms relying on graph decomposition \citep{Giakoumakis1997,Courcelle2000}.

The article is structured as follows. Section~\ref{sec:decomposition} considers the problem of decomposing a MaxClique instance into smaller subproblems which fit D-Wave. The efficiency of the decomposition algorithm relies on the computation of upper and lower bounds on the clique size of the subgraph it generates, and a variety of techniques for computing such bounds is presented in section~\ref{sec:pruning}, including subgraph extractions methods. Section~\ref{sec:simulations} contains an experimental analysis of the sensitivity of our decomposition approach on the chosen subgraph extraction techniques and on the bounds of the clique size, as well as an analysis of the effectiveness of these techniques for very sparse and very dense graphs compared to earlier work. The article concludes with a discussion in section~\ref{sec:discussion}.

\section{Maximum Clique Decomposition}
\label{sec:decomposition}
We use the CH-partitioning introduced in \cite{Djidjev2015}, see also \cite{Chapuis2017}, in order to split a large input graph $G=(V,E)$ into smaller subgraphs on which a maximum clique is found. We next briefly review the method.

\subsection{CH-partitioning}
\label{sec:CH}
\begin{figure}[t]
    \centering
    \includegraphics[width=0.5\textwidth]{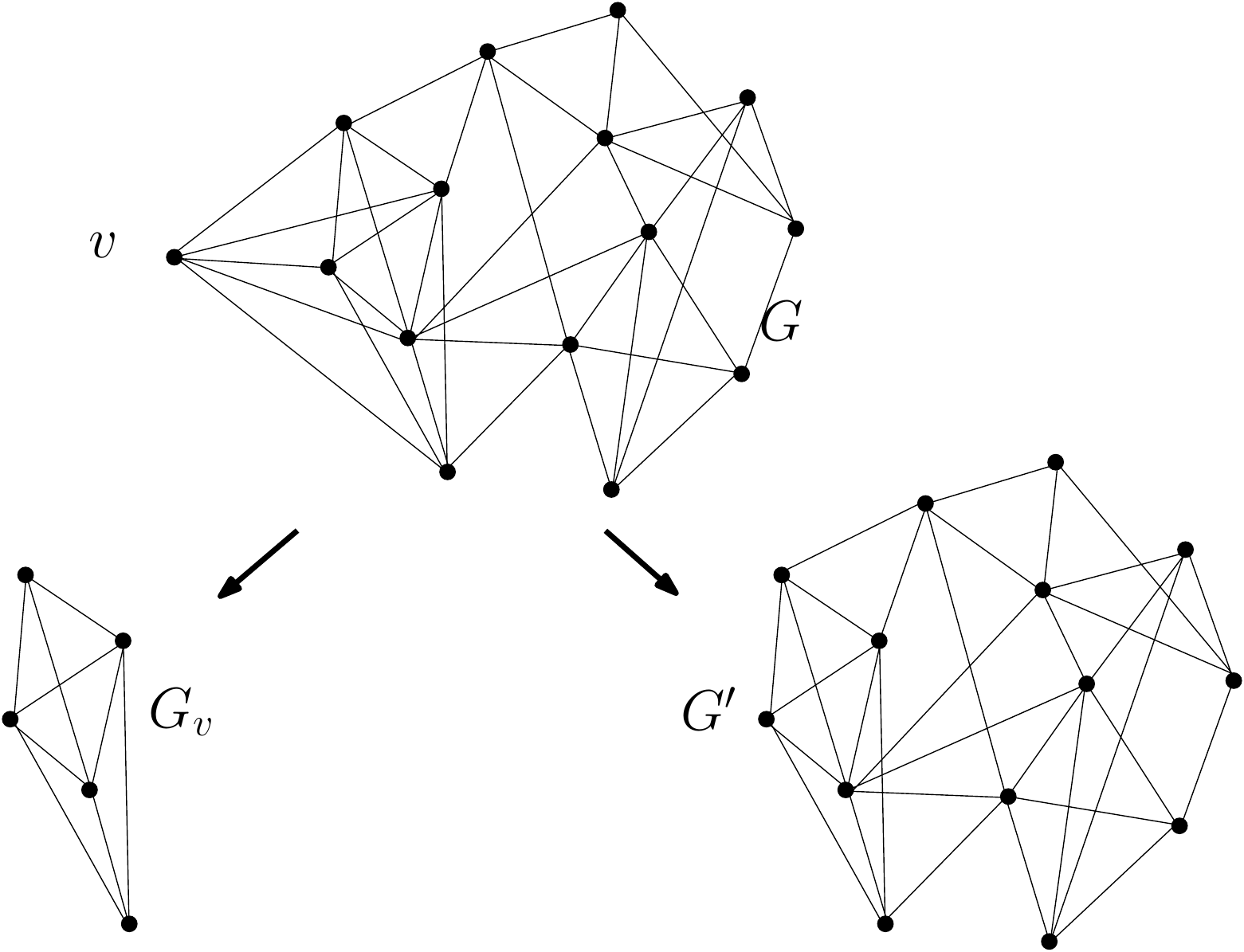}
    \caption{Illustration of the vertex splitting at a single vertex $v$.\label{fig:vertex_splitting}}
\end{figure}
The CH-partitioning algorithm works as follows. We iteratively select a single vertex $v \in V$ and extract the subgraph $G_v$ containing  all neighbors of $v$ and all edges between them. Afterwards, $v$ and all edges adjacent to $v$ are removed from $G$, thus creating another new graph denoted as $G'$. This is visualized in figure~\ref{fig:vertex_splitting}. The clique number $\omega(G)$ of $G$ is equal to $\min(\omega(G_v)+1,\omega(G'))$.

Before decomposing $G'$ and $G_v$ further in a similar fashion, our algorithm computes lower and upper bounds on the maximum clique sizes contained in $G'$ and $G_v$. This is helpful for two reasons. First, if the current best lower bound is less (worse) than the lower bound of a newly generated subgraph, we can update the current best lower bound accordingly. Second, if the upper bound for $G'$ or $G_v$ is less than the best current lower bound for the maximum clique, we can prune that subgraph and the branch of the decomposition tree associated with it. Our aim is to prune as many generated subproblems as possible, and section~\ref{sec:pruning} highlights the approaches we explore.

The algorithm stops splitting any generated subgraph as soon as it can be embedded on the D-Wave hardware, meaning if the subgraph size is at most $46$ ($65$) vertices in case of D-Wave 2X (D-Wave 2000Q).  The solution time of problems that fit the D-Wave hardware is essentially constant (anneals only without post-processing), so we are primarily concerned with the number of generated subproblems and the effectiveness of our pruning approaches. However, we do take into account that the success rate of annealing may depend on the graph structure and we compute different times-to-solutions for different graph types.

The proposed algorithm can be characterized as a branch-and-bound (B\&B) algorithm, a class of algorithms often used to compute a solution of NP-hard optimization problems. As such, B\&B has been previously used in exact classical algorithms for solving MaxClique, e.g., \cite{PARDALOS1992363,Carmo2012}. However, these algorithms explore the search space in a quite different way than the one studied in this paper, as they work bottom up, starting with a single-vertex clique and expanding it as much as possible. In comparison, our algorithm is top-down, starting with the original graph and splitting it down to smaller graphs, which are examined for a maximum clique. Such structure allows for the generation of a set of smaller size MaxClique problems which fit in and are solvable on the quantum processing unit directly. 

\subsection{Vertex Choice}
\label{sec:vertexchoice-description}
The algorithm of section~\ref{sec:CH} offers one more tuning parameter: the procedure for selecting the vertex $v$ that is used in each iteration to split the current graph instance $G$ into two new graphs. Possible choices include:
\begin{enumerate}[label=(\roman*),wide]
    \item a vertex $v$ of lowest degree.
    \item a vertex $v$ of median degree.
    \item a vertex $v$ chosen at random.
    \item a vertex $v$ of highest degree.
    \item a vertex $v$ in the $k$-core of lowest degree. The $k$-core of a graph $G = (V, E)$ is a maximal subgraph of $G$ in which every vertex has a degree of at least $k$. Specifically, we iteratively extract the $k$-core of sizes $k \in \{1,\ldots,|G|\}$ and stop at the first $k$ leading to a $k$-core which is not the original graph. We then choose $v$ as a vertex which was removed in that $k$-core reduction step.
    \item a lowest degree vertex $v$ whose extracted subgraph $G_v$, defined in section~\ref{sec:CH}, is of lowest density.
\end{enumerate}
In any of the above cases, if multiple vertices satisfy the selection criterion, the vertex $v$ which is extracted is chosen at random. The above vertex selection approaches are experimentally explored in section~\ref{sec:vertexchoice}.

\section{Pruning of subproblems}
\label{sec:pruning}
Next we describe two pruning approaches we use in order to reduce the number of subgraphs generated in the algorithm of section~\ref{sec:decomposition}. Section~\ref{sec:bounds} describes bounds which can be used to prune entire subgraphs. Section~\ref{sec:reduction} describes reduction approaches which allow to reduce the size of generated subgraphs, and sometimes entirely remove subgraphs, before decomposing them further.

\subsection{Bounds}
\label{sec:bounds}
We consider two approaches to compute upper bounds, and two approaches to compute lower bounds, on the maximum clique size $\omega(G)$ of each subgraph $G$ generated in the graph decomposition algorithm of section~\ref{sec:decomposition}. As upper bounds we consider the following:
\begin{enumerate}[label=(\roman*),wide]
    \item We use a greedy search heuristic to find an upper bound on the chromatic number of the graph. The \textit{chromatic number} of a graph $G = (V, E)$, denoted as $\chi(G)$, is the minimum number of colors needed to color each vertex of $G$ such that no edge connects two vertices of the same color. Since each vertex in a clique must have a distinct color we know that $\chi(G) \geq \omega(G)$ \citep{Knuth1993TheTheorem}, thus the chromatic number of $G$ is an upper bound on $\omega(G)$. Computing the chromatic number is NP-hard, so its exact computation would be intractable, but there are much better heuristics for its approximation compared with the ones for the clique number. Therefore, a greedy search heuristic for the chromatic number could provide an easily computable upper bound for the maximum clique number. In the simulations of section~\ref{sec:simulations}, we use two greedy algorithms: the greedy coloring algorithm \textit{greedy\_color} of the python module \textit{Networkx}~\citep{Networkx} as well as the upper bound algorithm of \cite[eq.~(4)]{Budinich2003ExactGraph} for graphs with densities larger than $0.8$.
    \smallskip
    \item The \textit{Lov\'asz number} of the complement of a graph $G$, denoted as $\theta(\overline{G})$, is defined in \citep{Lovasz1979} as an upper bound on the Shannon capacity of $\overline{G}$, and is sandwiched in-between the chromatic number $\chi(G)$ and the clique number $\omega(G)$, thus $\chi(G) \geq \theta(\overline{G}) \geq \omega(G)$ \citep{Knuth1993TheTheorem}. Therefore, $\omega(\overline{G})$ might improve upon the bound already given by $\chi(G)$, and cannot be worse. Although the Lov\'asz number can be calculated in polynomial time \citep{Knuth1993TheTheorem}, existing algorithms are not very scalable (due to large constants) since they involve solving a semidefinite programming problem. The python implementation to compute the Lov\'asz number, adapted from \citep{Stahlke2013}, becomes prohibitively slow for larger graphs: we only use it for subgraphs $|G| \leq 60$, and otherwise revert to the greedy search heuristic outlined above.
\end{enumerate}

Likewise, we consider the following two lower bounds:
\begin{enumerate}[label=(\roman*),wide]
    \item Any maximum clique heuristic can be used to provide a lower bound on the clique number, provided the result returned by the heuristic is a clique. We use the \textit{Fast Maximum Clique Finder} of \cite{fmcjournal,fmc} in heuristic mode.
    \smallskip
    \item At any point in the decomposition tree, a lower bound on one of the current subbranches can be obtained simply by solving the other subbranch. This gives a lower bound for the maximum clique size in all nodes (and leaves) of the yet unsolved branch.
\end{enumerate}

It should be noted that there are many other easily computable upper and lower bounds for maximum clique, maximum independent set, and chromatic number that generally apply to any type of graph \citep{Soto2011ThreeNumber, Elphick2018Conjectured2018}. However, in experiments we conducted (not reported in this article) those generally applicable bounds proved to be quite conservative, with the exception of the bound of \cite[eq.~(4)]{Budinich2003ExactGraph} which we employ for high densities.

\subsection{Reduction algorithms}
\label{sec:reduction}
We implement three types of reduction algorithms. The first two work directly on the subgraphs, the last one works with the QUBO formulation of MaxClique. Due to the inherent similarity between the first two techniques, we group them together as one reduction algorithm.
\begin{enumerate}[label=(\roman*),wide]
    \item The \textit{(vertex) $k$-core algorithm} can reduce the number of vertices of the input graph in some cases, and the \textit{edge $k$-core algorithm} \citep{Chapuis2017, DBLP:journals/corr/cs-DS-0310049} can reduce the number of edges.

    The (vertex) $k$-core of a graph $G = (V, E)$ was defined in section~\ref{sec:vertexchoice-description} as the maximal subgraph of $G$ in which every vertex has a degree of at least $k$. Therefore, if a graph has a clique $C$ of size $k+1$, then this clique $C$ must be contained in the $k$-core of $G$ and all vertices outside of the $k$-core can be removed.

    The edge $k$-core of a graph $G$ is defined in \cite{Chapuis2017}. It is easily shown that for two vertices $v$, $w$ in a clique of size $c$, the intersection $N(v) \cap N(w)$ of the two neighbor lists $N(v)$ and $N(w)$ of $v$ and $w$ has size at least $c-1$. Denoting the current best lower bound on the clique size as $L$ we can therefore choose a random vertex $v$ and remove all edges $(v,e)$ satisfying $|N(v) \cap N(e)| < L-2$, since such edges cannot be part of a clique with size larger than $L$.
    \smallskip
    \item A second approach aims to extract information from the QUBO formulation of MaxClique in \eqref{eq:MC_QUBO}. For a general QUBO \eqref{eq:hamiltonian}, persistency analysis of \citep{Boros2002} allows us to identify the values that certain variables must take in any minimum 
    (called \textit{strong persistencies}) or in at least one minimum (called \textit{weak persistencies}), as well as relations among them. We apply \eqref{eq:MC_QUBO} to any newly generated subgraph, compute a persistency analysis for the resulting QUBO and remove a vertex $v$ from the subgraph if its corresponding variable $x_v \in \{0,1\}$ (indicating if $v$ belongs to the maximum clique or not) could be allocated a value. Furthermore, the count of solved variables which are guaranteed to be in the maximum clique for a particular subgraph can improve the lower bound on the clique number for the overall algorithm.
\end{enumerate}

We use the vertex and edge $k$-cores algorithms with $k$ being set to the current best lower bound value for the clique number. This allows us to prune entire subgraphs that cannot contain a clique of size larger than the best current one. The persistency analysis allows to remove single vertices and their adjacent edges from subgraphs, as well as improve the lower bound on the clique number.

\section{Experimental analysis}
\label{sec:simulations}

\begin{table}[t]
\centering
    \begin{tabular}{|l|| l|l|l|l|l|l|l|l|l|}
        \hline
        density & 0.1 & 0.2 & 0.3 & 0.4 & 0.5 & 0.6 & 0.7 & 0.8 & 0.9\\
        \hline\hline
        p & 0.007 & 0.003 & 0.008 & 0.015 & 0.020 & 0.003 & 0.003 & 0.007 & 0.012\\
        $t_\text{run}$ & 1.938 & 2.057 & 2.317 & 3.382 & 6.736 & 6.981 & 4.975 & 4.333 & 3.429\\
        TTS & 0.127 & 0.296 & 0.141 & 0.105 & 0.152 & 0.948 & 0.746 & 0.273 & 0.137\\
        \hline
    \end{tabular}
    \caption{TTS times as a function of the intended graph density (first row). The time $t_\text{run}$ is the sum of averages for embedding ($T_\text{embedding}$) and QPU solve time ($T_\text{QPU}$).\label{table:dwave_tts}}
\end{table}

We investigate the effectiveness of the vertex choice of section~\ref{sec:vertexchoice-description} and the pruning techniques of section~\ref{sec:pruning}, and we compare an algorithm combining both to existing work. For each graph density in the interval $[0.1,0.9]$, incremented in steps of $0.1$, we generate random graphs of size varying from $65-110$ vertices and decompose them using only the partitioning outlined in \ref{sec:CH}. We then generate random graphs based on each subgraph's density and size, and solve them on D-Wave. As a performance measure we report the \textit{Time-To-Solution} measure (defined below) and/or the number of subgraphs generated, both as a function of the graph density. The Time-To-Solution (TTS) measure is defined in \cite[eq.~(3)]{Mandra2018} as
\begin{equation}
	\text{TTS} = (T_\text{embedding} + T_\text{QPU}) \frac{\log(0.01)}{\log(1-p)},
	\label{eq:TTS}
\end{equation}
where $T_\text{embedding}$ and $T_\text{QPU}$ are the average times in seconds (for $10^4$ anneals per graph) to compute an embedding onto the D-Wave architecture and to perform annealing, and $p$ is proportion of anneals which correctly found the maximum clique.

Table~\ref{table:dwave_tts} shows the proportion of anneals which found the maximum clique on D-Wave, the average times $t_\text{run}$ and the resulting TTS times for each of the graph densities considered. We fix the TTS times given in table~\ref{table:dwave_tts} and use them to report runtimes in the remainder of section~\ref{sec:simulations}.

As test data, we use instances of $G(n,p)$ random graphs \citep{ErdosRenyi1960} generated with density $p$ and fixed graph size $n=80$, which the exception of figure~\ref{table:dimacs} which uses selected DIMACS benchmark graphs \citep{dimacs}. All results are averaged over $20$ repetitions unless otherwise stated.

To calculate TTS for each DIMACS graph tested, we decomposed the graph being considered until we had less than or equal to 500 subgraphs. We then randomly selected 60 of those subgraphs as representatives, determined their sizes and densities, and for each we solved a random subgraph of similar size and density on D-Wave. We then applied eq.~\ref{eq:TTS} to the resulting data.

\subsection{Vertex Choice}
\label{sec:vertexchoice}
\begin{figure}[t]
    \centering
    \includegraphics[width=0.49\textwidth]{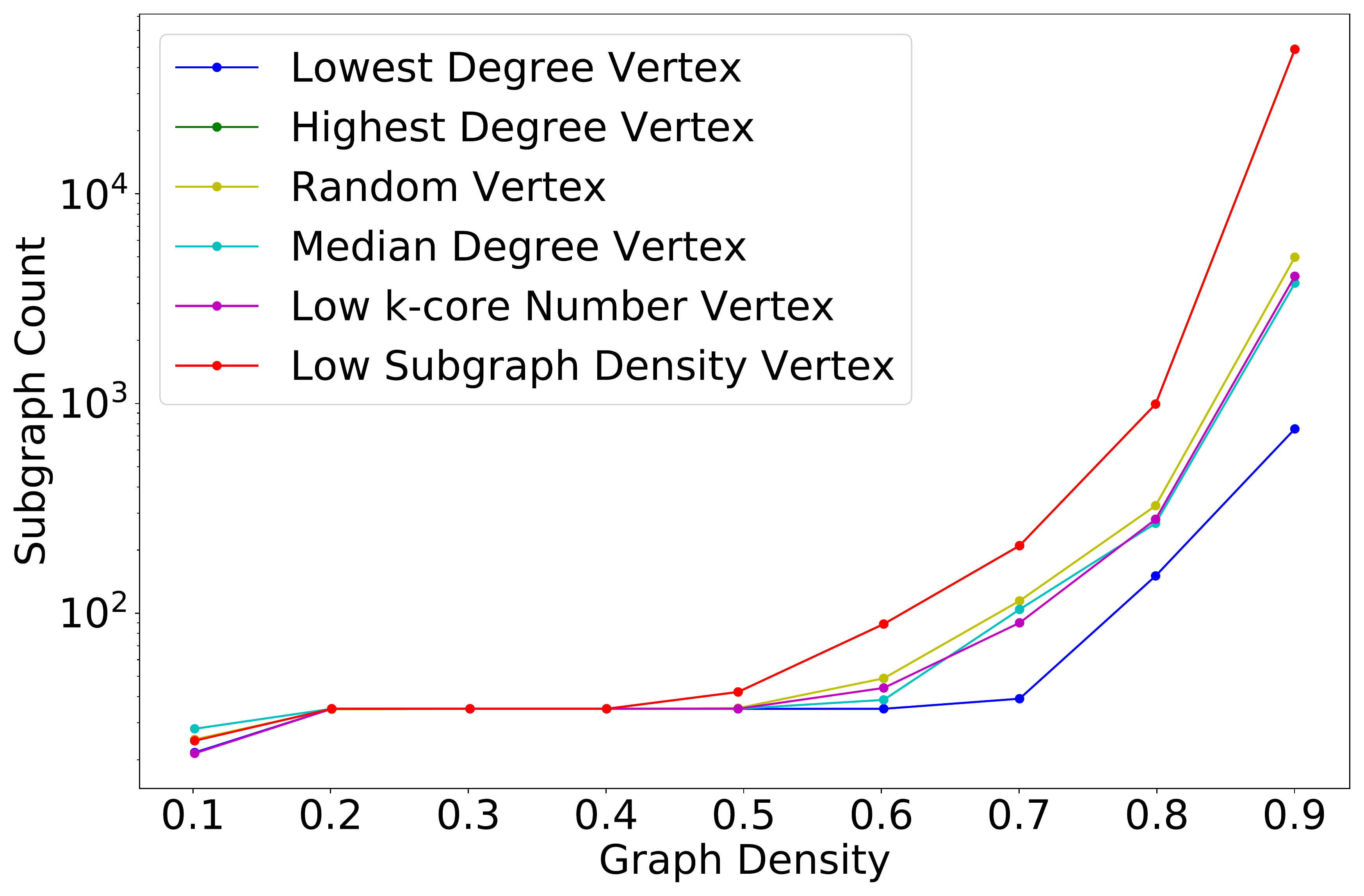}~
    \includegraphics[width=0.49\textwidth]{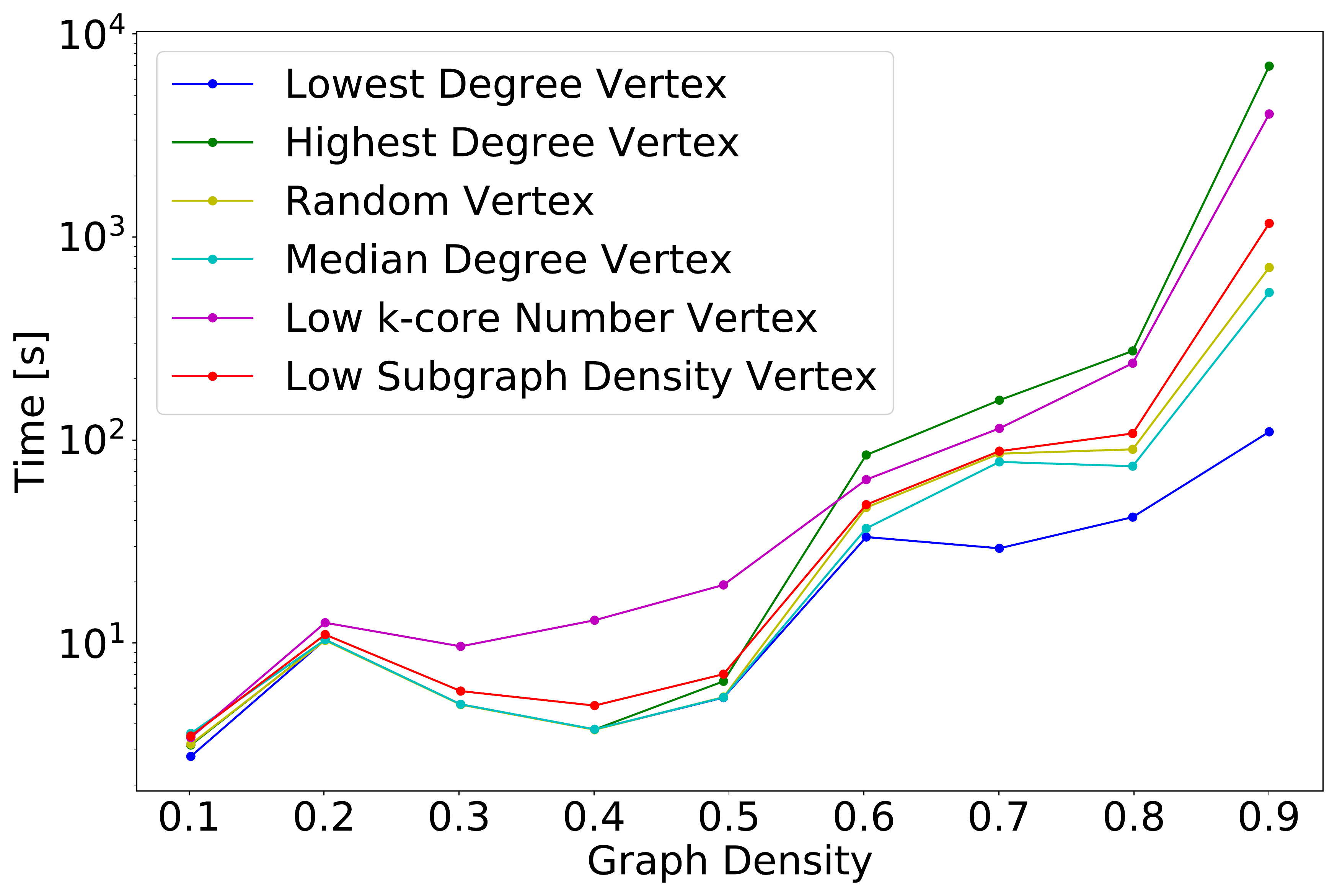}
    \caption{Subgraph count (left) and total solution time (right) as a function of the graph density for the six vertex selection approaches of section~\ref{sec:vertexchoice-description}. Log scale on the y-axes.\label{fig:vertex_choice}}
\end{figure}
Figure~\ref{fig:vertex_choice} shows the total solution time to decompose our test graphs (to the point that each subproblem reaches a size of at most $46$ vertices, thus making it possible to compute the maximum clique on D-Wave) as a function of the graph density. We compare the six different vertex selection choices delineated in section~\ref{sec:vertexchoice-description}. The lowest degree selection strategy seems to lead to the best solution time among the six approaches, where the advantage of the lowest degree selection becomes more pronounced as the graph density increases. Similar results are observed for the number of subgraphs.

\subsection{Bounds}
\label{sec:boundcomparison}
\begin{figure}[t]
    \centering
    \includegraphics[width=0.49\textwidth]{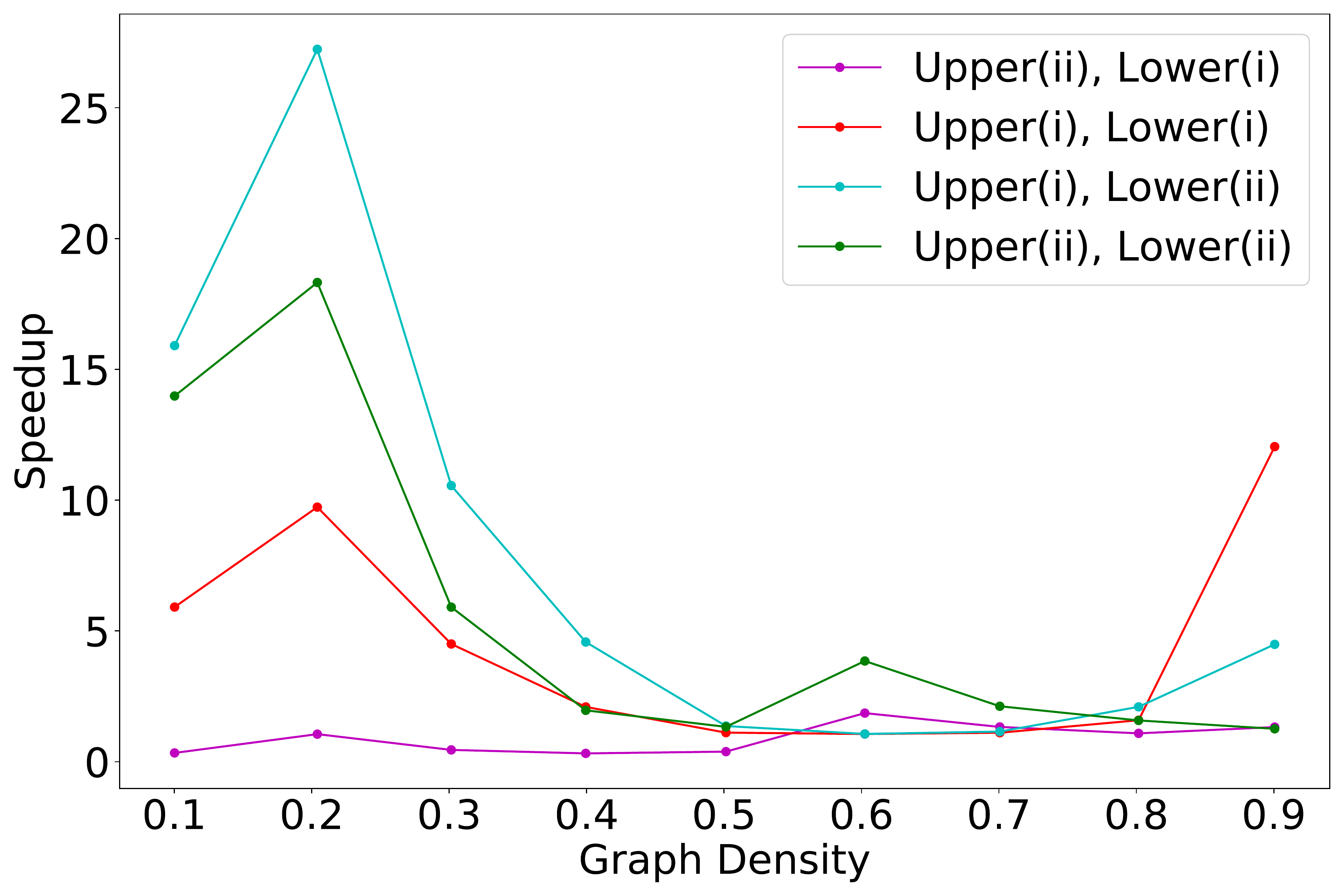}~
    \includegraphics[width=0.49\textwidth]{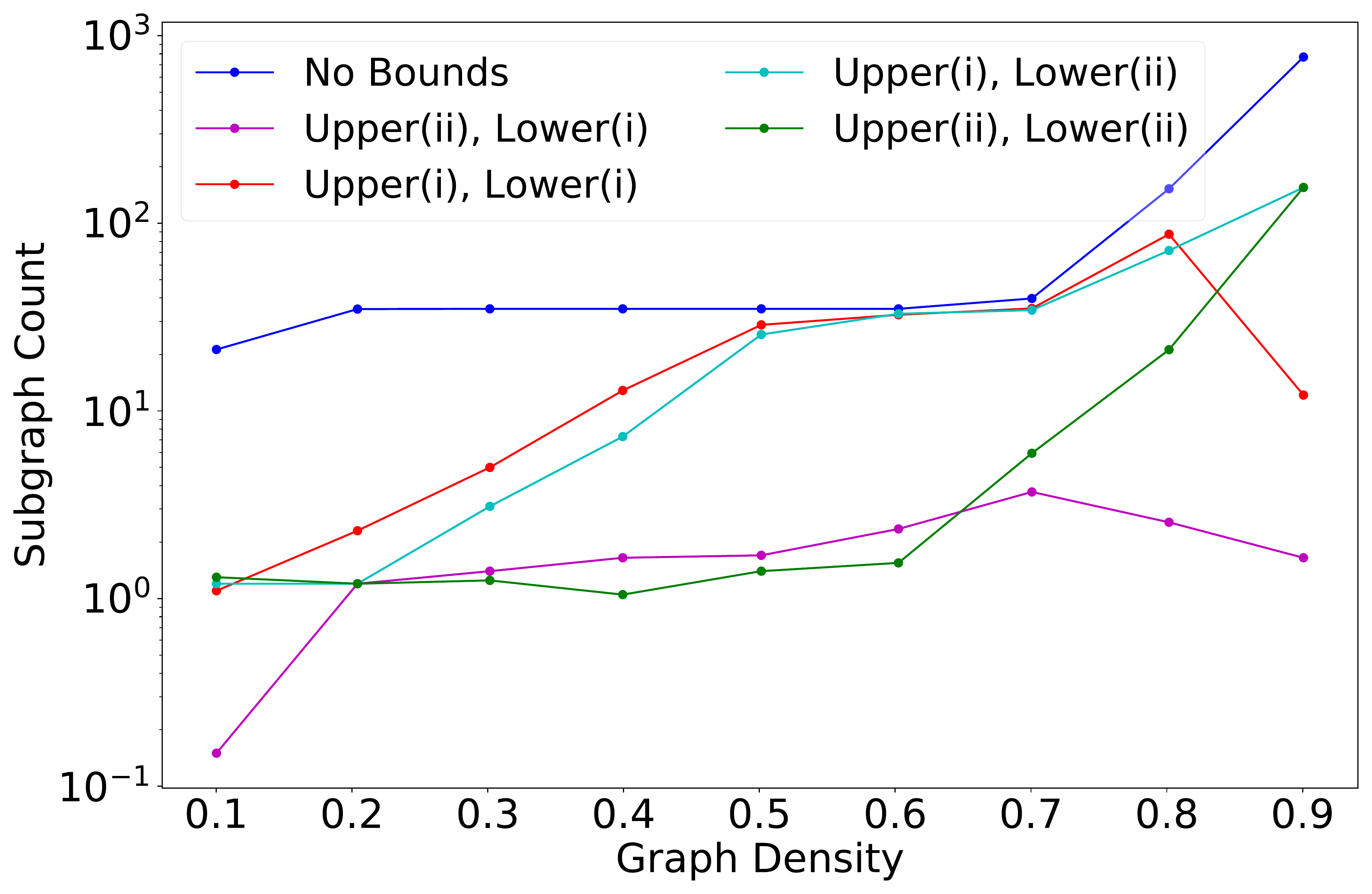}
    \caption{Subgraph count (left) and ratio $T_0/T_1$ (right) defined in section~\ref{sec:boundcomparison} as a function of the graph density for the two techniques to compute upper bounds on the clique number (see section~\ref{sec:bounds}), in comparison to no bound calculations. Log scale on the y-axis showing subgraph count.\label{fig:Bound_Solve_Time}}
\end{figure}

Figure~\ref{fig:Bound_Solve_Time} (left) shows the number of subgraphs generated by every combination of the two lower and upper bounds outlined in section~\ref{sec:bounds}. Figure~\ref{fig:Bound_Solve_Time} (right) shows the speedup in runtime when using the upper bounds of section~\ref{sec:bounds} compared to not using any bounds. The speedup is defined as $T_0/T_1$, where $T_0$ ($T_1$) is the total solution time without (with) bounds. The right figure shows that the Lov\'asz number bound is not an efficient bound in terms of solution time due to the  preprocessing CPU time required to calculate it. However, the Lov\'asz number leads to the lowest number of generated subgraphs (left figure). Over most densities tested, the combination of fast heuristic bounds given by upper bound (i) and lower bound (ii) (see section~\ref{sec:bounds}) seem to give a reasonable trade-off between total solution time and subgraph count. We will use this choice in section~\ref{sec:Comparisons}.

\subsection{Reduction algorithms}
\begin{figure}[t]
    \centering
    \includegraphics[width=0.49\textwidth]{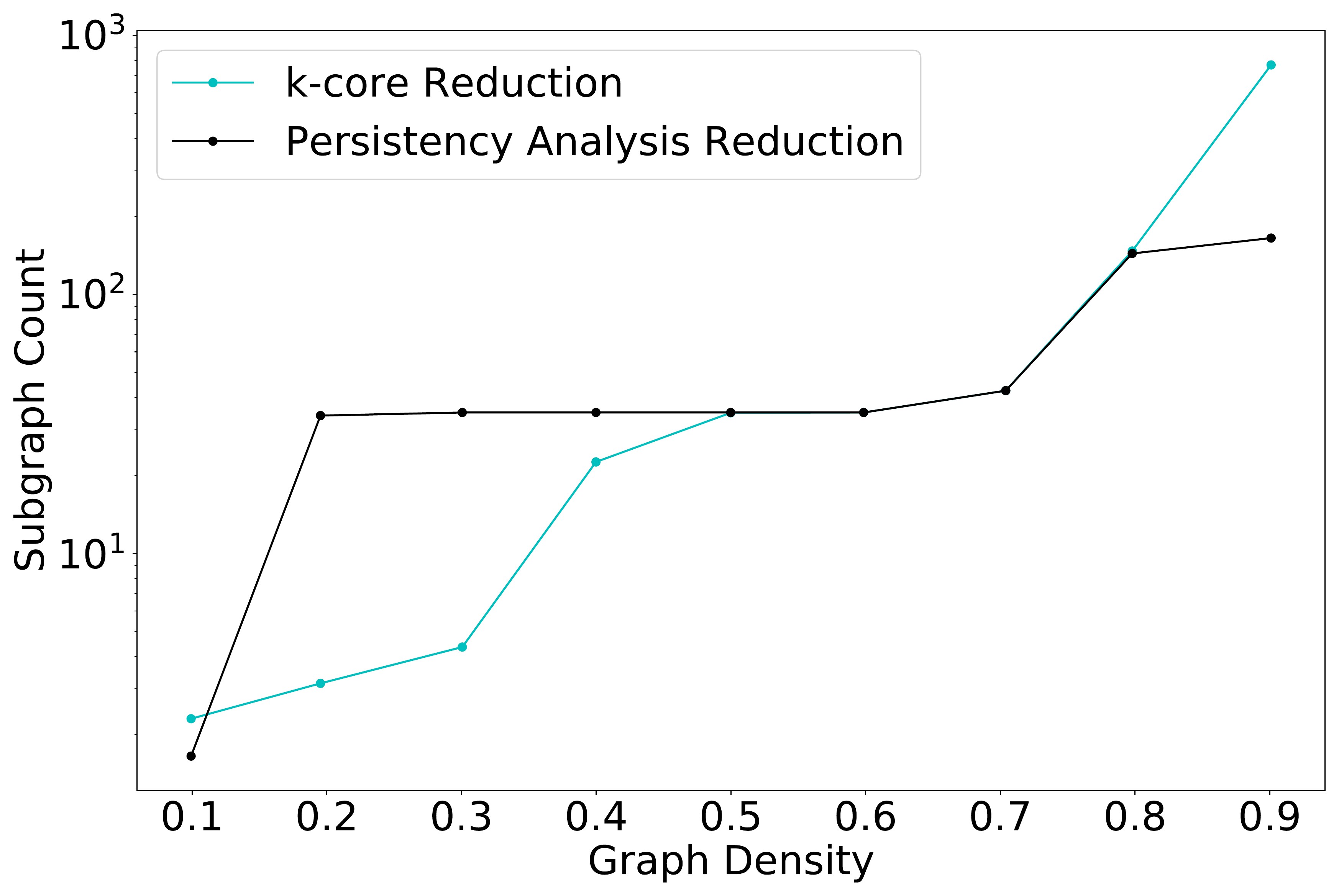}\hfill
    \includegraphics[width=0.49\textwidth]{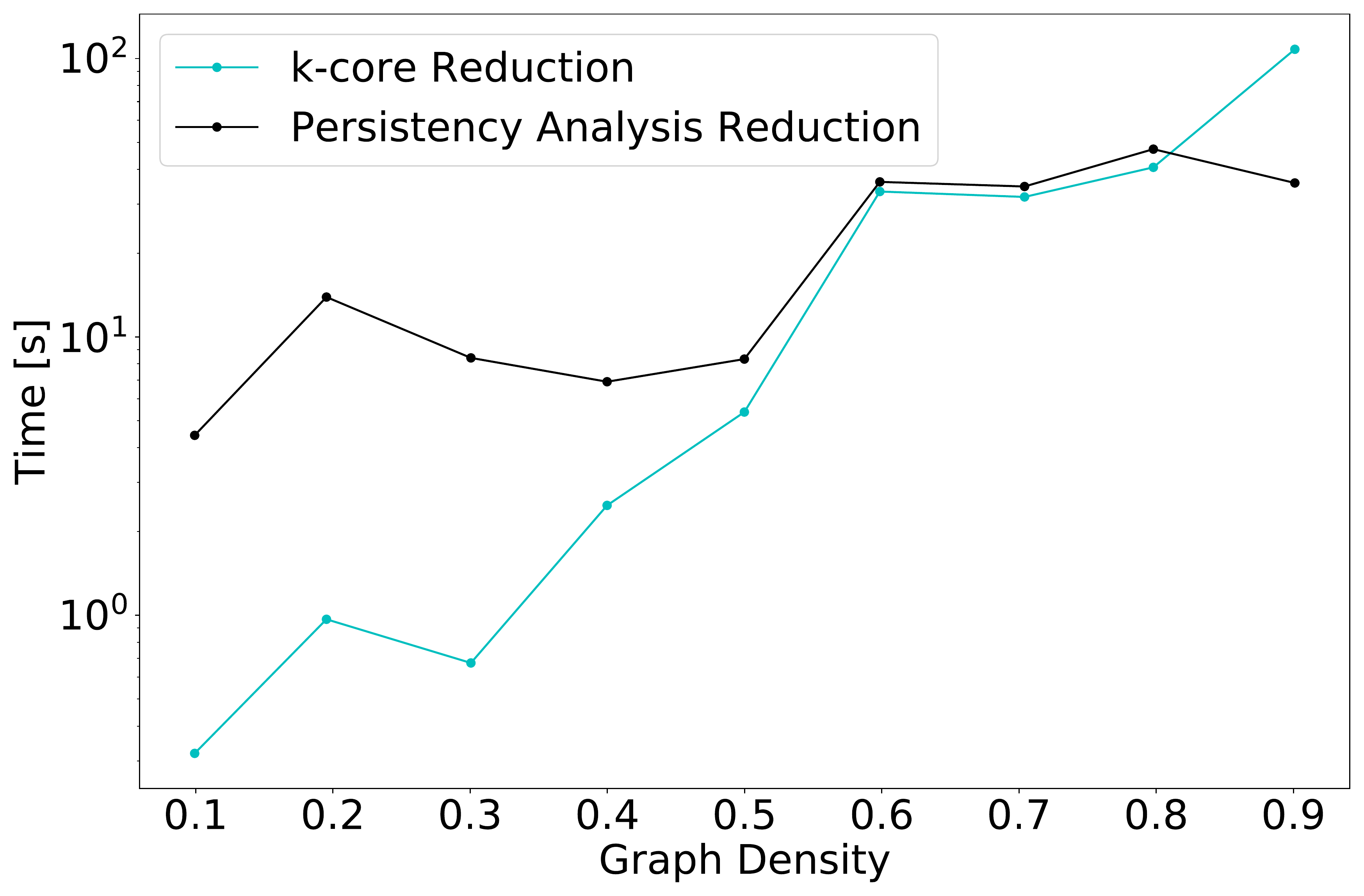}
    \caption{Subgraph count (left) and total solution time (right) for the two reduction techniques from section~\ref{sec:reduction}. Log scale on the y-axes.\label{fig:k-core}}
\end{figure}

Figure~\ref{fig:k-core} (right) shows the total runtime to decompose our test graphs as a function of the density $p$. Two methods are compared: the $k$-core reduction, and the persistency analysis (section~\ref{sec:reduction}). Figure~\ref{fig:k-core} (left) shows that the $k$-core reduction algorithm can significantly reduce the number of subgraphs, and therefore the total solution time, over more densities than the persistency analysis can, and that the latter yields improvements in runtime only for high densities with $p>0.8$ (figure~\ref{fig:k-core}, right).\\[2ex]

\subsection{Comparisons}\label{sec:Comparisons}
\begin{figure}[t]
    \centering
    \includegraphics[width=0.5\textwidth]{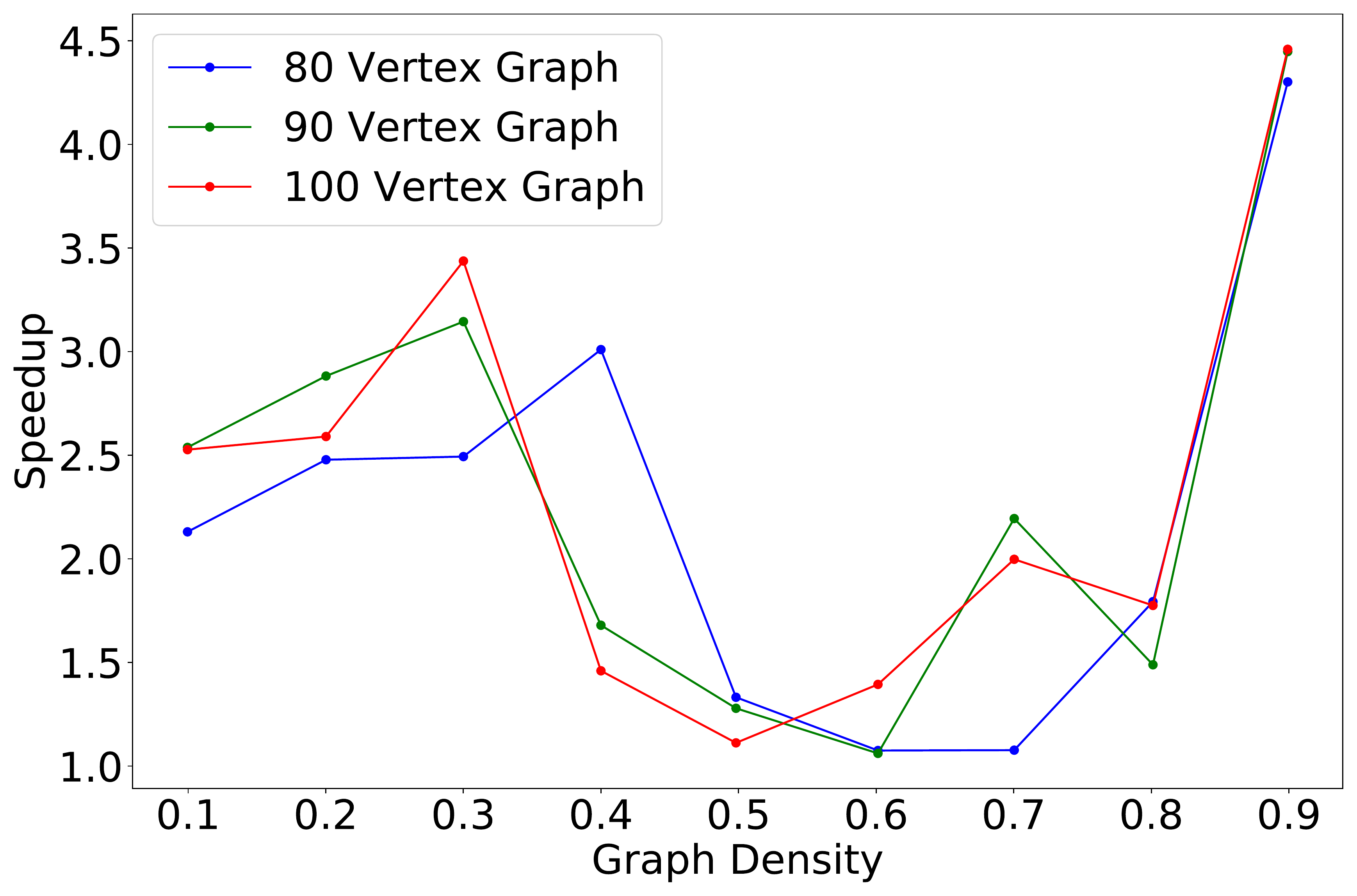}
    \caption{Ratio $T_\text{Chapuis}/T_\text{DBK}$ defined in section~\ref{sec:Comparisons} showing the speedup for the DBK algorithm in comparison to the algorithm of \cite{Chapuis2017} for low and high densities.\label{fig:comparison}}
\end{figure}

We denote the decomposition algorithm of section~\ref{sec:decomposition} which uses the choice of fast heuristic bounds of section~\ref{sec:bounds} for pruning subgraphs and the $k$-core of section~\ref{sec:reduction} as the DBK algorithm (Decomposition, Bounds, $k$-core). Figure~\ref{fig:comparison} shows the speedup of the DBK algorithm over the decomposition algorithm presented in \cite{Chapuis2017}. To be precise, figure~\ref{fig:comparison} displays $T_\text{Chapuis}/T_\text{DBK}$, where $T_\text{DBK}$ is the time the DBK algorithm takes for a particular graph instance, and analogously for $T_\text{Chapuis}$. The figure shows that both algorithms are quite similar for graph densities in the interval $[0.4,0.8]$, however DBK considerably improves upon \cite{Chapuis2017} the sparser or denser the graphs are.

For DIMACS graphs displayed in table~\ref{table:dimacs}, we observe that the DBK algorithm can yield substantial improvements in runtime especially for large graphs, and that it never performs worse than \cite{Chapuis2017}.
\begin{table}[t]
\centering
\begin{tabular}{|l||l|l|l||l|l|}
    \hline
    DIMACS graph ~ & No.~vertices~ & No.~edges~ & $\omega(G)$~ & DBK algorithm~ & Chapuis et al.~(2017)~\\
    \hline
    johnson16-2-4 & 120 & 5460 & 8 & 3217 & 3217 \\
    keller4 & 171 & 9435 & 11 & 682 & 1310 \\
    p-hat300-1 & 300 & 10933 & 8 & 16 & 20 \\
    p-hat300-2 & 300 & 21928 & 25 & 794 & 2781 \\
    p-hat500-1 & 500 & 4459 & 9 & 61 & 77 \\
    p-hat700-1 & 700 & 60999 & 11 & 175 & 643 \\
    brock200-2 & 200 & 9876 & 12 & 72 & 295 \\
    brock200-3 & 200 & 12048 & 15 & 1185 & 1640 \\
    brock200-4 & 200 & 13089 & 17 & 2168 & 14735 \\
    hamming6-2 & 64 & 1824 & 32 & 4 & 17 \\ 
    hamming8-4 & 256 & 20864 & 16 & 5179 & 18253 \\
    \hline
\end{tabular}
\caption{Total solution time in seconds for DIMACS graphs based on a single run.\label{table:dimacs}}
\end{table}

\section{Discussion}
\label{sec:discussion}
This article investigates subgraph extraction and pruning techniques for a decomposition algorithm designed to compute maximum cliques of a graph. These reduction techniques aim at reducing the number of recursively generated subgraphs for which the maximum clique problem needs to be solved. The overall runtime of our algorithm depends on the user's choice of method to solve the decomposed maximum clique subproblems at leaf level of the decomposition. Our analysis shows the following:
\begin{enumerate}
    \item The decomposition is sensitive to the vertex choice used to extract subgraphs. Extraction of the subgraph induced by the lowest degree vertex seems to lead to the lowest overall runtime.
    \item A variety of lower and upper bounds exist to estimate the maximal clique size and help prune subgraphs which cannot contain the maximum clique. Of those investigated, the Lov\'asz bound is most effective in pruning subgraphs but also most computationally intensive, whereas a heuristic upper bound leads to a good trade-off between subgraph reduction and runtime.
    \item The $k$-core reduction prunes subgraphs more effectively than an alternative approach based on persistency analysis.
    \item Combining the decomposition algorithm with subgraph extraction and fast heuristic bounds on the clique size for pruning leads to an improved algorithm which, despite the additional computations in each recursion, is faster at computing maximum cliques than a previous algorithm of \cite{Chapuis2017}, with a considerable speed-up for very sparse and very dense graphs.
\end{enumerate}

Future research could investigate the viability of trade-off algorithms, where differing decomposition and simplification strategies are used in response to different problem graph types. Another future improvement concerns the Lov\'asz number: our experiments show that its bounds yield the best reduction in subproblem count, but also come at the highest computational cost (see figure~\ref{fig:Bound_Solve_Time}). Additionally, the impracticality of applying the adapted python implementation \citep{Stahlke2013} on large graphs was not ideal. A faster way to compute the Lov\'asz number would render those bounds usable in practice.


\end{document}